# AAU-net: An Adaptive Attention U-net for Breast Lesions Segmentation in Ultrasound Images

Gongping Chen, Lei Li, Yu Dai, Jianxun Zhang, and Moi Hoon Yap

***Abstract*—Various deep learning methods have been proposed to segment breast lesions from ultrasound images. However, similar intensity distributions, variable tumor morphologies and blurred boundaries present challenges for breast lesions segmentation, especially for malignant tumors with irregular shapes. Considering the complexity of ultrasound images, we develop an adaptive attention U-net (AAU-net) to segment breast lesions automatically and stably from ultrasound images. Specifically, we introduce a hybrid adaptive attention module (HAAM), which mainly consists of a channel self-attention block and a spatial self-attention block, to replace the traditional convolution operation. Compared with the conventional convolution operation, the design of the hybrid adaptive attention module can help us capture more features under different receptive fields. Different from existing attention mechanisms, the HAAM module can guide the network to adaptively select more robust representation in channel and space dimensions to cope with more complex breast lesions segmentation. Extensive experiments with several state-of-the-art deep learning segmentation methods on three public breast ultrasound datasets show that our method has better performance on breast lesions segmentation. Furthermore, robustness analysis and external experiments demonstrate that our proposed AAU-net has better generalization performance in the breast lesion segmentation. Moreover, the HAAM module can be flexibly applied to existing network frameworks. The source code is available on https://github.com/CGPxy/AAU-net**

## I. Introduction

Breast cancer is a common female disease, which seriously threatens women's health and life [1], [2]. Therefore, regular breast screening and diagnosis are very important to formulate treatment plans and improve survival rates. Due to the flexibility and convenience of ultrasound imaging, it has become a convention modality for breast tumors screening [3]. The segmentation of breast ultrasound (BUS) images can help us characterize and localize breast tumors, which is one of the key steps in computer-aided diagnosis (CAD) [4]. In recent years, many deep learning methods have been proposed to segment breast lesions from ultrasound images [5]. However, complex ultrasound patterns, similar intensity distributions, variable tumor morphologies and blurred boundaries seriously interfere with the segmentation accuracy of breast lesions, and even breast tumors cannot be detected, as shown in Fig. 1.

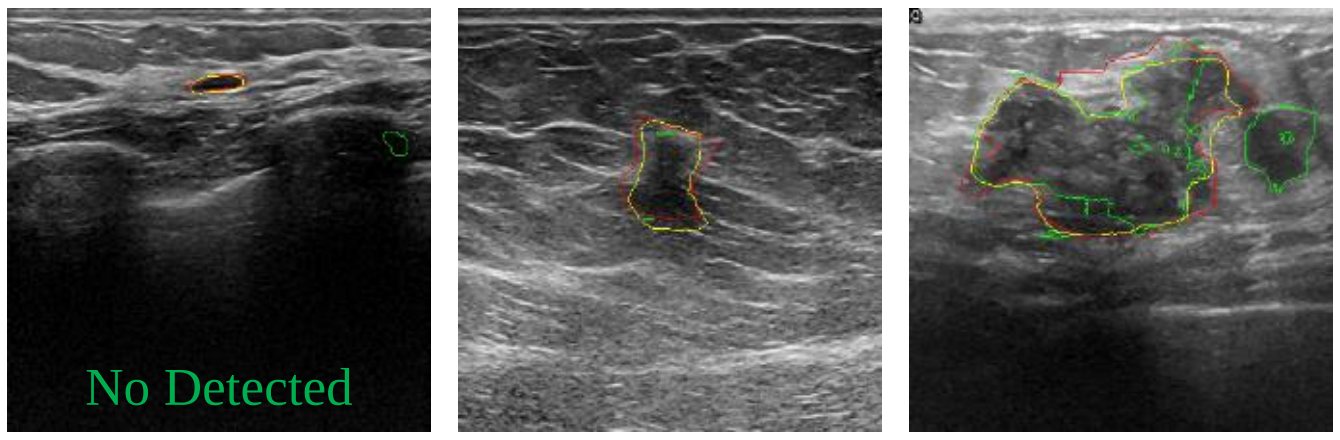

Fig. 1. Various BUS images and the segmentation results by U-net and our method. The red curve is the ground-truth boundary of the lesion. The yellow and green curves are the segmentation results of our method and U-net, respectively. It can be seen from these images that tumor morphology, blurred borders and similar surrounding tissue (background) severely affect the segmentation accuracy of breast lesions, especially for small and malignant tumors.

The powerful nonlinear learning ability makes full convolution network (FCN) and U-net have achieved great success in medical images segmentation [2], [6]–[13]. Enlightened by this, many deep learning methods are proposed to segment breast lesions from ultrasound images [5], [14]–[17]. In 2018, Yap *et al.* are the first to systematically evaluate the impact of different FCN variants on breast lesions segmentation and achieve segmentation results that outperform traditional methods [18]. In 2018, Almajalid *et al.* improved the segmentation accuracy of U-net for breast lesions by contrast enhancement and speckle noise removal strategy [19]. These preprocessing operations can improve network performance, but they destroy the original spatial feature distribution of objects. Moreover, Ning *et al.* pointed out that the complex ultrasound pattern and the interference of surrounding tissue make it difficult for these simple frameworks to achieve ideal segmentation results on BUS images [20], as shown in Fig. 1.

To further refine the segmentation results of BUS lesions, two optimization strategies: enlarging the receptive field [10], [21]–[23] and the attention mechanism [24]–[27] have been widely used. The dilated convolution operation is a commonly used strategy to expand the receptive field [28]–[30]. For example, Hu *et al.* obtained the large receptive field of breast tumors by using dilated convolutions in deeper network layers

This work is supported by the National Natural Science Foundation of China Grant U1913207 and Grant 51875394. (Corresponding author: Yu Dai, e-mail: daiyu@nankai.edu.cn)

Gongping Chen, Yu Dai and Jianxun Zhang are with the College of Artificial Intelligence, Nankai University, Tianjin, 300350 China (e-mail: cgp110@mail.nankai.edu.cn, daiyu@nankai.edu.cn, zhangjx@nankai.edu.cn).

Lei Li is with Institute of Biomedical Engineering, University of Oxford, London, U.K. (e-mail: lei.li@eng.ox.ac.uk)

Moi Hoon Yap is with the School of Computing, Mathematics and Digital Technology, Manchester University, Manchester M1 5GD, U.K. (e-mail: m.yap@mmu.ac.uk).

[30]. However, Xue *et al.* point out that deeper convolutional layers tend to focus more on the extraction of local features, and it is difficult to obtain the true global view using dilated convolutional operations on deeper convolutional layers [27]. Irfan *et al.* developed an end-to-end semantic segmentation network using dilated convolution operations to segment breast tumors from the BUS image [28]. Cao *et al.* integrated a set of hybrid dilated convolutions into $D^2$U-Net to alleviate the challenges posed by different lesion sizes and shapes [29]. However, merely using dilated convolution operations to obtain a larger receptive field cannot fully cope with the perturbation caused by surrounding tissues and blurred boundaries [31].

In terms of attention mechanism, Lee *et al.* proposed a channel attention module to further improve the performance of U-net for breast lesions segmentation [26]. Abraham *et al.* utilized an optimized attention U-net, which includes three modules of multi-scale inputs, attention U-net and deep supervision, to segment breast tumors from ultrasound images [24]. Since U-net uses $3\times3$ convolution operations, the attention mechanism can only be performed on the fixed receptive field, which limits the segmentation performance of the U-net variant network [32]. Inspired by the work of Li *et al.* [32], Byra *et al.* constructed U-net with selected kernel convolution (SKNet) to segment breast lesions [31]. However, the original selection kernel convolution ignores feature selection under different receptive fields in spatial dimensions. Recently, Xue *et al.* developed a global guidance network for breast lesions segmentation by integrating a channel attention module, a spatial attention module, and a boundary detection module [27]. Huang *et al.* proposed a novel boundary-rendering framework to further refine the contours of breast lesions [2]. These two methods further alleviate the interference of various factors in the segmentation accuracy of breast lesions.

To obtain the segmentation results of breast lesions closer to the ground-truth mask, some segmentation networks integrating dilated convolution and attention mechanisms are proposed [33], [34]. Yan *et al.* proposed an attention enhanced U-net (AE U-net) with hybrid dilated convolution to segment breast tumors from ultrasound images [33]. Zhuang *et al.* developed a residual dilated attention-gate U-net (RDAU-Net) to segment breast lesions by introducing dilated convolution and residual learning strategies on attention U-net [34]. Although the performance of segmentation networks can be optimized by integrating dilated convolution and attention mechanisms, there are still some shortcomings (fixed receptive field sizes and single-attention operations) that need to be further overcome.

To alleviate the above challenge, we design a hybrid adaptive module to aggregate information from multiple kernels to achieve adaptive receptive field size of neurons. The designed hybrid adaptive module mainly consists of three parts: convolutional layers with different kernel sizes, channel self-attention block, and spatial self-attention block. Different from existing methods, the hybrid adaptive module can adaptively select receptive fields with different scales from the channel and spatial dimensions. Subsequently, we use the hybrid adaptive module to propose a novel adaptive attention U-net to improve breast lesions segmentation by learning generic representations from BUS images. Compared with conventional convolution operations with fixed receptive fields, the hybrid adaptive attention module can help the network select more robust representations from multiple perspectives. Extensive experiments demonstrate that our adaptive attention U-net brings significant and consistent improvements in breast lesions segmentation. Our main contributions are as follows:

- We design a novel hybrid adaptive attention module, which can adaptively select receptive fields with different scales from the channel and spatial dimensions.
- A novel adaptive attention U-net is developed to segment breast lesions from ultrasound images. The network can improve breast lesions segmentation by learning generic representations from BUS images.
- Extensive experiments on three public BUS datasets demonstrate that our approach consistently improves the segmentation accuracy of breast lesions, outperforming the strong baseline and the state-of-the-art medical image segmentation methods.

In the remainders of this paper, we introduce our segmentation network and loss function in Section II. The public ultrasound datasets, experimental settings and evaluation indicators are elaborated in Section III. The experimental results are presented in Section IV. Finally, our discussion and conclusions are given in Sections V and VI, respectively.

## II. Methodology

The Fig. 2 is an illustration of our developed adaptive attention U-net (AAU-net) for breast lesions segmentation. Our AAU-net has the same core architecture as U-net [10] mainly including four down-sampling, four up-sampling and four skip-connections. The difference is we propose a hybrid adaptive attention module (HAAM) to replace the original convolution layer to better adapt to the segmentation of breast lesions. As shown in Fig. 2, each encoding or decoding stage consists of two HAAMs. Convolutional layers with different kernel sizes in HAAM can provide receptive fields with different scales, which can improve the adaptability of the network to different input images. Moreover, AAU-net can learn more robust representations from BUS images through channel dimension and spatial dimension constraints.

### *A. Hybrid Adaptive Attention Module (HAAM)*

The designed HAAM mainly consists of three parts: convolutional layers with different kernel sizes, channel self-attention block, and spatial self-attention block. Specifically, the input HAAM feature maps are processed by three parallel convolutional layers to obtain three feature maps with different receptive fields. The three convolutional layers are $3\times3$ convolution, $5\times5$ convolution, and $3\times3$ dilated convolution with a dilation rate of $3$. The feature map captured by three convolution layers can be represented as:

$$F^3 = W_{3\times3} \times F_{input} \tag{1}$$

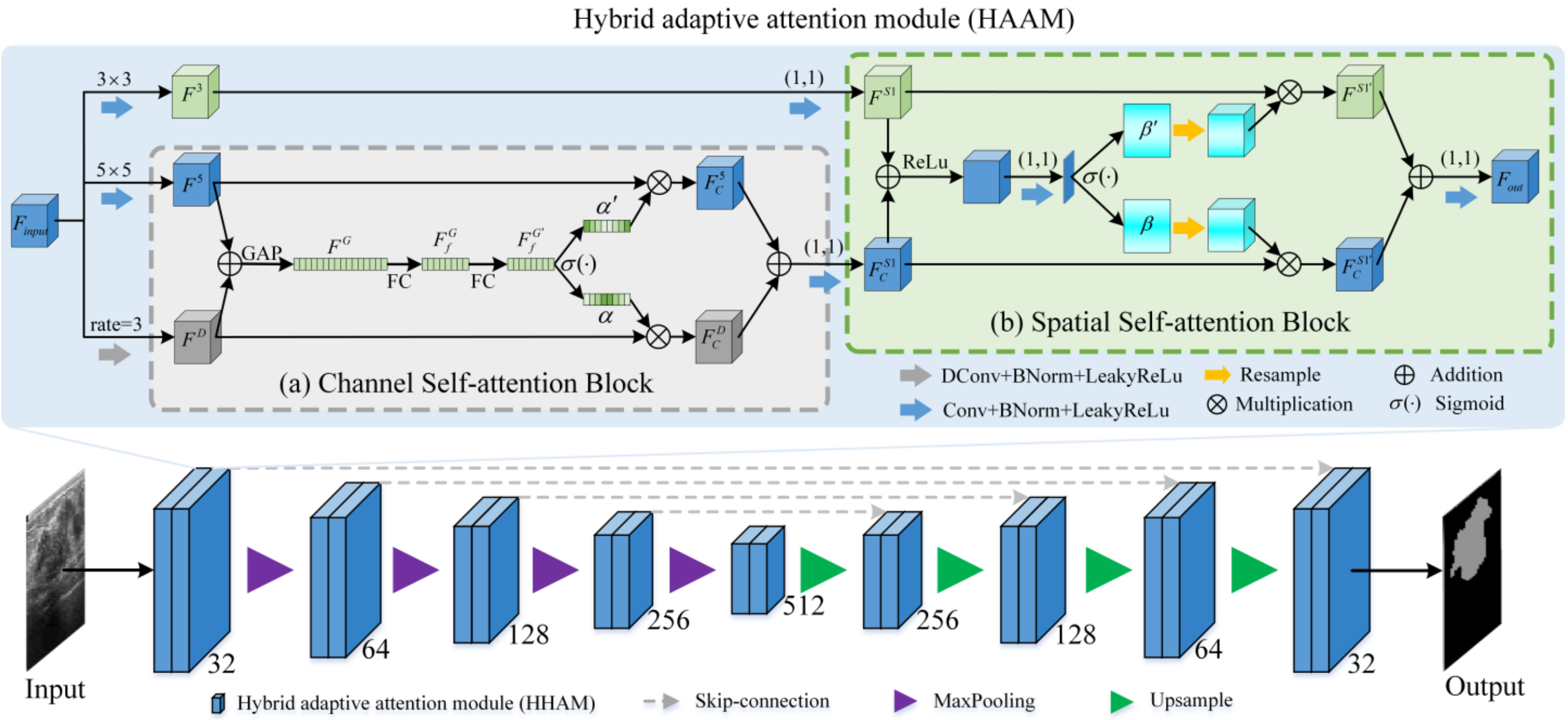

Fig. 2. The description of the adaptive attention U-net (AAU-net). The network is still a U-shaped network including four down-sampling and four up-sampling operations. Each stage consists of two hybrid adaptive attention modules (HAAM).

$$F^5 = W_{5\times5} \cdot F_{input} \tag{2}$$

$$F^D = W_{3\times3}^D \cdot F_{input} \tag{3}$$

where $F_{input} \in \mathbb{R}^{c' \times h \times w}$ represents the input feature map, $W_{3\times3}$ and $W_{5\times5}$ denote the matrix of $3\times3$ convolution and $5\times5$ convolution, respectively. $W_{3\times3}^D$ represents the matrix of dilated convolution. $F^3 \in \mathbb{R}^{c \times h \times w}$ and $F^5 \in \mathbb{R}^{c \times h \times w}$ denote the feature map captured by $3\times3$ convolution layer and $5\times5$ convolution layer, respectively. $F^D \in \mathbb{R}^{c \times h \times w}$ indicates the feature map captured by dilated convolution layer. The receptive field sizes captured by the three convolutional layers are shown in Fig. 3. Subsequently, these feature maps are integrated into channel self-attention block and spatial self-attention block. It can be seen from Fig. 3 that capturing receptive fields with different scales from input images can not only improve the adaptability of the network to different inputs, but also better characterize BUS images.

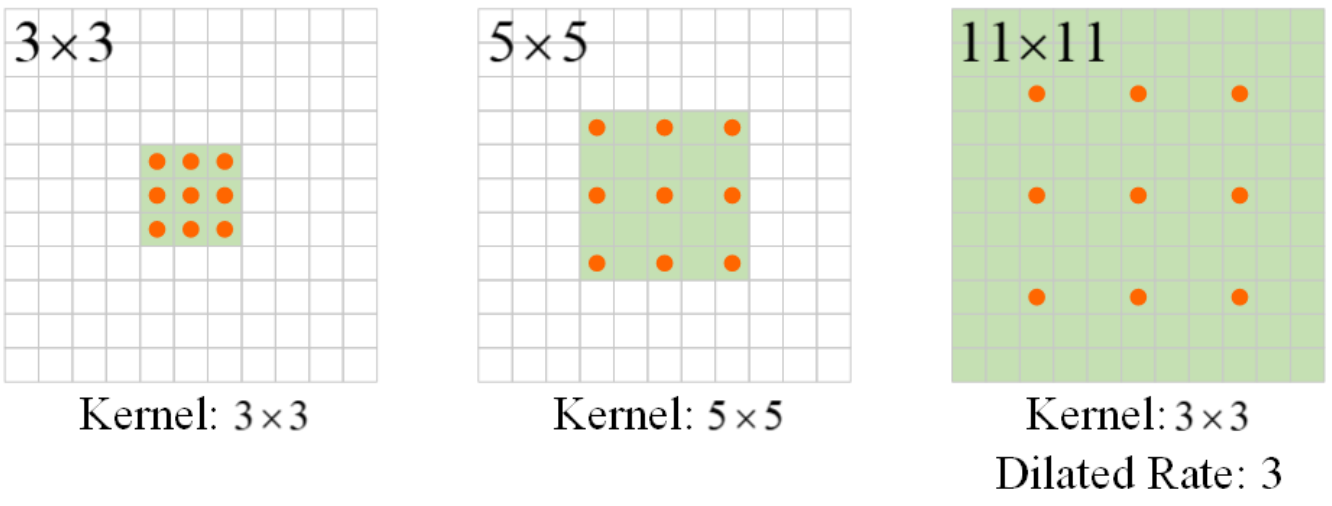

Fig. 3. Receptive fields captured by three convolutional layers. We can see that the receptive field obtained by $5\times5$ convolution layer is equivalent to 2 convolution layers with kernel size is $3\times3$, and the receptive field of dilated convolutions is the same as that of 5 convolution layers with kernel size is $3\times3$.

## *B. Channel Self-Attention Block*

To capture useful objective features from different receptive fields, we first develop a channel self-attention module to adaptively guide the network to learn more robust feature representations. Fig. 2 (a) illustrates the channel self-attention block, which purpose is to guide the segmentation network to select more representative features from the channel dimension. Specifically, we first compress the combined feature maps of $F^5 \in \mathbb{R}^{c \times h \times w}$ and $F^D \in \mathbb{R}^{c \times h \times w}$ into a new feature map $F^G \in \mathbb{R}^{2c \times 1 \times 1}$ of size $1\times1$ through a global average pooling (GAP) operation. The obtained feature can be expressed as:

$$F^G = GAP(F^5 \oplus F^D) \tag{4}$$

where $F^5$ and $F^D$ denote the feature map captured by $5\times5$ convolution layer and dilated convolution layer, respectively. $\oplus$ represents the element-wise addition. Then, the feature map $F^G$ is input to a fully connected layer followed by a batch-normalization layer and a ReLU layer to produce a new feature map. The obtained feature can be expressed as:

$$F_f^G = \sigma_r(B(W_{fc} \cdot F^G)) \tag{5}$$

where $W_{fc}$ represents the matrix of fully connected layers. $B(\cdot)$ and $\sigma_r(\cdot)$ denote batch-normalization and ReLU activation operations, respectively. The feature map $F_f^G$ again undergoes a fully connected operation to obtain a new feature map,

$$F_f^{G'} = W_{fc} \cdot F_f^G \tag{6}$$

Finally, the feature map $F_f^{G'}$ is performed sigmoid activation to obtain the channel attention map:

$$\alpha = \sigma_s(F_f^{G'}) \tag{7}$$

We let $\alpha \in [0,1]^{c\times1\times1}$ and $\alpha' \in [0,1]^{c\times1\times1}$ represent the channel attention maps of $F^D$ and $F^5$, respectively. Each value of $\alpha/\alpha'$ indicates the importance of channel information at the corresponding voxel in $F^D/F^5$. It is worth noting that $\alpha'$ is derived from $\alpha$, and its value is $1-\alpha$. These two channel attention maps can help us adaptively extract more representative feature maps from receptive fields with different scales. In order to achieve automatic feature selection, we use the channel attention map $\alpha$ to calibrate the feature map $F^D$, and the channel attention map $\alpha'$ to calibrate the feature map $F^5$. The feature map after channel attention map calibration can be expressed as:

$$F_C^D = \alpha \otimes F^D \tag{8}$$

$$F_C^5 = \alpha' \otimes F^5 \tag{9}$$

Subsequently, the feature maps $F_C^D \in \mathbb{R}^{c\times h\times w}$ and $F_C^5 \in \mathbb{R}^{c\times h\times w}$ are integrated and used as the input of the next stage.

### C. Spatial Self-Attention Block

As we all know, the channel attention mechanism focuses on the category of the feature, and the spatial attention mechanism focuses on the location of the feature [35]. To further improve the robustness of network representation features, we develop a novel spatial self-attention block as shown in Fig. 2 (b). The feature maps obtained by the $3\times3$ convolution layer and the channel self-attention block are used as the input of the spatial self-attention block. To refine the location information of the objective, we first perform a $1\times1$ convolution operation on the input feature map. The feature map after $1\times1$ convolution can be defined as:

$$F^{S1} = W_{1\times1} \cdot F^3 \tag{10}$$

$$F_C^{S1} = W_{1\times1} \cdot (F_C^5 \oplus F_C^D) \tag{11}$$

where $F_C^D$ and $F_C^5$ denote the output of the channel self-attention block, $F^3$ represents the feature map obtained by performing $3\times3$ convolution operations on the input of HAAM. Subsequently, the feature maps fused with $F^{S1}$ and $F_C^{S1}$ undergo a ReLU activation $\sigma_r(\cdot)$, a $1\times1$ convolution operation and a sigmoid activation $\sigma(\cdot)$, to obtain the spatial attention map:

$$\beta = \sigma(W_{1\times1} \cdot \sigma_r(F^{S1} \oplus F_C^{S1})) \tag{12}$$

We let $\beta \in [0,1]$ and $\beta' \in [0,1]$ represent the spatial attention maps of $F_C^{S1}$ and $F^{S1}$, respectively. It is worth noting that the value of $\beta'$ is $1-\beta$. Each value of $\beta/\beta'$ indicates the importance of spatial information at the corresponding voxel in $F_C^{S1}/F^{S1}$. To perform calibration on $F_C^{S1}$, $\beta$ is resampled to obtain a spatial attention map with the same number of channels as $F_C^{S1}$. Similarly, the resample operation is performed for $\beta'$. The feature maps calibrated by $\beta$ and $\beta'$ can be denoted as $F_C^{S1'}$ and $F^{S1'}$, respectively. Finally, the output of the spatial self-attention block is obtained after the connected $F_C^{S1'}$ and $F^{S1'}$ are subjected to the convolution operation.

$$F_{out} = W_{1\times1} \cdot (F_C^{S1'} \oplus F^{S1'}) \tag{13}$$

$F_{out}$ is also the output of the entire hybrid adaptive attention module (HAAM).

### D. Loss Function

Binary cross entropy (BCE) [36] is one of the widely used loss functions in two-class image segmentation tasks, which reflects the direct difference between predicted masks and ground-truth labels. Its definition can be expressed as:

$$\ell_{BCE} = -\sum_{(i,j)} Y(i,j)\cdot\log\hat{Y}(i,j) + (1-Y(i,j))\cdot\log(1-\hat{Y}(i,j) \tag{14}$$

where $Y(i,j)\in[0,1]$ denotes the ground-truth label of the pixel $(i,j)$, $\hat{Y}(i,j)\in[0,1]$ represents the predict masks. In this study, we use BCE loss for the training of the network.

TABLE I
SAMPLE DISTRIBUTION OF THE THREE PUBLIC BUS DATASET.

| | Benign | Malignant | Normal | Total | Cross-validation | External-validation |
|---|---|---|---|---|---|---|
| BUSI | 437 | 210 | 133 | 780 | True | False |
| Dataset B | 110 | 53 | No | 163 | True | False |
| STU | Unknow | Unknow | No | 42 | False | True |

## III. DATASETS AND EXPERIMENTAL SETTINGS

### A. BUS Datasets

In this paper, three widely used public BUS datasets with different scales are used to evaluate the segmentation network performance. Table I describes the sample distribution of these three public BUS datasets. The first BUS dataset (denotes as BUSI) is constructed by Al-Dhabyani *et al.* [37]. The dataset contains 780 images acquired by two types of ultrasound equipment (LOGIQ E9 ultrasound and LOGIQ E9 Agile ultrasound system) in the Baheya Hospital. The average image size of these images is $500\times500$ pixels. The second BUS dataset used in this paper named Dataset B is collected by Yap *et al.* [38]. Dataset B contains 163 images with average image size of $760\times570$ pixels collected by Siemens ACUSON Sequoia C512 system. The third public BUS dataset is the STU provided by Zhuang *et al.* [34]. The STU contains 42 BUS images with average image size of $128\times128$ pixels acquired by the Imaging Department of the First Affiliated Hospital of Shantou University using the GE Voluson E10 ultrasonic diagnostic system. Since the STU dataset contains too few images, it is only used as external validation data to evaluate the generalization performance of the segmentation network.

TABLE II
THE SEGMENTATION RESULTS (MEAN ± STD) OF DIFFERENT NETWORK COMPONENTS ON BUSI AND DATASET B. RED ARROWS REPRESENT INCREASES. THE BEST RESULTS ARE MARKED WITH BOLD TEXTS.

| | | Jaccard | Precision | Recall | Specificity | Dice |
|---|---|---|---|---|---|---|
| BUSI | Baseline U-net | 60.70±2.36 | 71.88±2.41 | 76.30±2.48 | 96.18±0.55 | 70.10±2.20 |
| | U-net with channel self-attention block | 62.43±1.95↑ | 74.63±1.21↑ | 75.64±1.83↑ | 97.13±0.85↑ | 72.06±1.04↑ |
| | U-net with spatial self-attention block | 65.12±1.10↑ | 76.11±1.43↑ | 78.02±1.21↑ | 97.45±0.96↑ | 75.86±1.07↑ |
| | U-net with HAAM (Ours) | **68.82±0.44**↑ | **79.61±1.07**↑ | **80.10±0.52**↑ | **97.57±0.24**↑ | **77.51±0.68**↑ |
| Dataset B | Baseline U-net | 58.44±4.26 | 70.27±6.11 | 75.32±2.85 | 98.44±0.40 | 68.20±4.23 |
| | U-net with channel self-attention block | 62.76±3.60↑ | 74.28±4.56↑ | 78.45±4.11↑ | 98.74±0.28↑ | 72.37±3.29↑ |
| | U-net with spatial self-attention block | 66.94±2.26↑ | 76.33±2.61↑ | 80.97±4.03↑ | 98.75±0.39↑ | 75.77±1.84↑ |
| | U-net with HAAM (Ours) | **69.10±2.98**↑ | **78.83±2.40**↑ | **82.22±3.84**↑ | **98.82±0.35**↑ | **78.14±2.41**↑ |

### B. Experimental Settings

To fully verify the effectiveness and robustness of our method, we use three datasets to conduct extensive experiments, such as ablation study, comparison with state-of-the-art segmentation methods, and robustness analysis. Our ablation studies mainly consist of component ablation and parameter ablation. In the ablation study, we perform four-fold cross-validation on BUSI and Dataset B, respectively. In comparative experiments with state-of-the-art segmentation methods, we perform four-fold cross-validation on BUSI and Dataset B, respectively. The robustness analysis mainly includes four parts: the robustness on benign and malignant lesions segmentation, the external validation, the comparison on BUSI with normal images and the comparison with different attention-based methods. Dataset B and STU contain too few malignant lesions, so we choose BUSI for robustness analysis of malignant lesions segmentation. Similarly, benign lesions in BUSI are selected to evaluate the robustness of different networks for segmenting benign lesions. Depending on the number of samples, we perform four-fold cross-validation on benign images and three-fold cross-validation on malignant images. In the external validation experiments, the STU dataset is used as test data to evaluate the segmentation performance of each method after training on Dataset B. We train each segmentation method by four-fold cross-validation on BUSI with normal images to evaluate the impact of normal ultrasound images in the breast lesion segmentation. In the comparative experiment with attention-based methods, we perform four-fold cross-validation on BUSI and Dataset B, respectively. During the training process, the training data and test data of each fold do not have any overlap.

We choose Adam optimizer to train our network. The initial learning rate of our network is 0.001. Multiple cross-validation show that the best segmentation performance is obtained when epoch size and batch size are set to 50 and 12, respectively. Our experimental device is a PC with two NVIDIA RTX 3090 GPUs. The development environment is Ubuntu 20.04, python 3.6 and TensorFlow 2.6.0.

### C. Evaluation Metrics

To quantitatively evaluate the segmentation performance of different methods on breast lesions, we use nine widely used segmentation metrics. For a detailed description of the nine evaluation indicators of Jaccard, Precision, Recall, Specificity, Dice, AUC, Hausdorff distance (HD), average boundary distance (ABD) and average symmetric surface distance (ASSD), please refer to [27], [39]. Due to the complexity of ultrasound patterns, existing deep learning segmentation methods are prone to fail to detect objective regions on individual images, as shown in Fig. 1 and Fig. 5. It is well known that boundary-based metrics cannot fairly evaluate these segmentation-failed images. To ensure the absolute fairness of the comparison, the three metrics (HD, ABD and ASSD) are only used in external validation experiments.

## IV. EXPERIMENTAL RESULTS

In this section, we first conduct the ablation study on the components and parameters of our network. Then, we compare our method with state-of-the-art deep learning segmentation methods. Finally, the robustness of our network is analyzed.

### A. Ablation Study

#### 1) Architecture Ablation

To evaluate the performance of different network components, we perform ablation experiments on BUSI and Dataset B. In the ablation experiments, U-net is used as the benchmark network and four-fold cross-validation is performed on BUSI and Dataset B, respectively. Table II shows the experimental results of different components on BUSI and Dataset B. The results of ablation experiments indicate that these network components designed in this paper all play a role in improving network performance. From Table II, we can see that the hybrid adaptive attention module (HAAM) enables the network to achieve the best segmentation results on BUSI and Dataset B. This suggests that integrating the constraints of channel self-attention block and spatial self-attention block can help the network learn more robust representations from BUS images.

#### 2) Parameter Ablation

To further evaluate the perturbation of receptive field size on segmentation performance, we analyze the impact of smaller and larger receptive fields on segmentation results. To generate the smaller receptive field, we first replace the 5×5 convolution with the 3×3 convolution, and then reduce the dilation rate of the 3×3 dilated convolution from 3 to 2. To obtain the larger receptive field, a 5×5 convolution is used to substitute the 3×3 convolution, and the remaining two convolutions do not make any changes. The comparison results of different receptive field sizes on BUSI and Dataset B are shown in Table III. According to the segmentation results in Table III, we can conclude that the smaller and larger receptive fields are not beneficial for breast lesions segmentation. The above comparison also proves the rationality of our network convolution kernel size and dilation rate settings.

TABLE III
THE SEGMENTATION RESULTS (MEAN ± STD) OF DIFFERENT PARAMETERS ON BUSI AND DATASET B. THE BEST RESULTS ARE MARKED WITH BOLD TEXTS.

| | Kernel size and Dilation rate | Jaccard | Precision | Recall | Specificity | Dice |
|---|---|---|---|---|---|---|
| BUSI | 3×3 ; 3×3 ; 3×3, $r = 2$ (the smaller receptive fields) | 68.15±1.02 | 79.02±1.69 | 80.59±0.83 | 97.17±0.93 | 77.05±0.95 |
| | 3×3 ; 5×5 ; 3×3, r=3 (Ours) | **68.82±0.44** | **79.61±1.07** | **81.10±0.52** | **97.57±0.24** | **77.51±0.68** |
| | 5×5 ; 5×5 ; 3×3, r=3 (the larger receptive fields) | 68.26±0.88 | 79.18±1.62 | 80.66±0.97 | 97.28±0.66 | 77.13±0.98 |
| Dataset B | 3×3 ; 3×3 ; 3×3, r=2 (the smaller receptive fields) | 68.32±3.93 | 78.17±3.26 | 81.64±4.13 | 98.33±0.35 | 77.52±2.49 |
| | 3×3 ; 5×5 ; 3×3, r=3 (Ours) | **69.10±2.98** | **78.83±2.40** | **82.22±3.84** | **98.82±0.35** | **78.14±2.41** |
| | 5×5 ; 5×5 ; 3×3, r=3 (the larger receptive fields) | 68.53±2.43 | 78.45±2.57 | 81.75±3.96 | 98.49±0.37 | 77.68±2.01 |

TABLE IV
THE SEGMENTATION RESULTS (MEAN ± STD) OF DIFFERENT COMPETING METHODS ON BUSI AND DATASET B. THE BEST RESULTS ARE MARKED WITH BOLD TEXTS. ASTERISKS INDICATE THAT THE DIFFERENCE BETWEEN OUR METHOD AND THE COMPETING METHOD IS SIGNIFICANT USING A PAIRED STUDENT'S T-TEST. (*: $p < 0.05$ ).

| | Method | U-net | Att U-net | RDAU-Net | U-net++ | Abraham *et al.* | U-net3+ | SegNet | AE U-net | SKNet | Ours |
|---|---|---|---|---|---|---|---|---|---|---|---|
| BUSI | Jaccard | 60.70±2.36 | 57.09±1.22 | 63.75±3.36 | 61.38±1.73 | 61.62±2.69 | 63.03±2.79 | 67.31±1.87 | 64.57±2.91 | 68.10±1.63* | **68.82±0.44** |
| | Precision | 71.88±2.41 | 78.78±4.67 | 71.25±4.11 | 79.68±3.07 | 73.77±2.90 | 71.89±3.28 | 76.09±2.00 | 74.44±3.74 | 78.62±1.66* | **79.61±1.07** |
| | Recall | 76.30±2.48 | 66.97±4.08 | 78.90±1.35 | 71.44±2.77 | 76.87±2.58 | 80.58±2.48 | 80.85±1.03* | 79.00±2.11 | 79.53±1.93 | **81.10±0.52** |
| | Specificity | 96.18±0.55 | 96.87±0.83 | 96.63±0.76 | 97.04±0.54 | 96.40±0.62 | 96.19±0.68 | 96.99±0.53 | 96.80±0.54 | 97.33±0.45* | **97.57±0.24** |
| | Dice | 70.10±2.20 | 67.99±1.18 | 71.94±3.46 | 71.58±2.09 | 71.35±2.67 | 71.85±2.73 | 75.64±1.80 | 73.47±3.03 | 76.92±1.57* | **77.51±0.68** |
| Dataset B | Jaccard | 58.44±4.26 | 59.93±4.53 | 58.17±4.91 | 61.19±5.86 | 63.09±3.04 | 65.63±5.26* | 62.83±2.20 | 62.37±2.16 | 64.25±4.01 | **69.10±2.98** |
| | Precision | 70.27±6.11 | 70.40±6.05 | 70.49±4.26 | 68.32±5.73 | 73.70±5.08 | 73.50±6.21 | 71.72±1.70 | 72.27±1.91 | 75.27±6.70* | **78.83±2.40** |
| | Recall | 75.32±2.85 | 76.15±4.21 | 73.55±5.28 | 79.64±3.84 | 79.24±1.72 | 80.29±3.93* | 80.15±3.90 | 78.97±2.29 | 79.36±2.50 | **82.22±3.84** |
| | Specificity | 98.44±0.40 | 98.43±0.33 | 98.37±0.39 | 98.44±0.41 | 98.61±0.36 | 98.60±0.36 | 98.59±0.30 | 98.67±0.28 | 98.68±0.39* | **98.82±0.35** |
| | Dice | 68.20±4.23 | 69.30±4.07 | 68.22±4.94 | 69.77±5.30 | 72.32±3.14 | 73.98±4.72* | 72.16±1.52 | 72.23±2.14 | 73.53±4.05 | **78.14±2.41** |

### B. Comparison with State-of-the-Art Methods

To evaluate the robustness and effectiveness of the method proposed in this paper, our method is first compared with state-of-the-art deep learning methods for BUS images and medical images segmentation. Our comparative methods include U-net [10], SegNet [9], Att U-net [25], U-net++ [7], U-net3+ [8], Abraham *et al.* [24], SKNet [31], AE U-Net [33] and RDAU-Net [34]. To ensure the fairness of the comparison, we perform four-fold cross-validation on BUSI and Dataset B, respectively. The quantitative evaluation results of different segmentation methods are presented in Table IV. From Table IV, we can see that our method achieves the best results on five evaluation metrics. The five evaluation index values of our method on BUSI are 67.97, 78.66, 80.63, 97.75 and 77.21, respectively. Compared to the second results, these metrics are improved by 5.4%, 4.4%, 1.5%, 0.6% and 4.3%, respectively. Compared to the second results on Dataset B, our method improves these five metrics by 5.3%, 4.7%, 2.4%, 0.1% and 5.6%, respectively. To further demonstrate the advantages of our method, we perform paired student's t-test with the second results, and the p-value ( $p < 0.05$ ) indicates a significant difference between our method and the comparison methods. From the above analysis, it can be concluded that our method has a very superior performance in breast lesions segmentation.

We also illustrate the P-R curves and the ROC curves of different segmentation methods on BUSI and Dataset B in Fig. 4. The P-R curve represents the confidence level that the true positive and false positive classes are predicted correctly. The ROC curve represents the confidence level that a method predicts correctly. The AUC scores are shown in the ROC curves. Compared to other methods, our method achieves the highest AUC values on both BUSI and Dataset B. According to the comparison of P-R and ROC curves, it can be concluded that our method achieves the highest confidence level on BUSI and Dataset B segmentation.

Fig. 5 shows the visual segmentation results on BUSI and Dataset B by different segmentation methods. Compared with the segmentation results of other methods, our method not only effectively alleviates the perturbation of tumor size, surrounding tissue and cascade, but also achieves segmentation results that are closer to the ground-truth masks. Moreover, the method proposed in this paper can alleviate the influence of heterostructure on segmentation results. Comprehensive evaluation results and visual effects show that our method achieves the best segmentation results with fewer missed and false detections in the breast lesion segmentation.

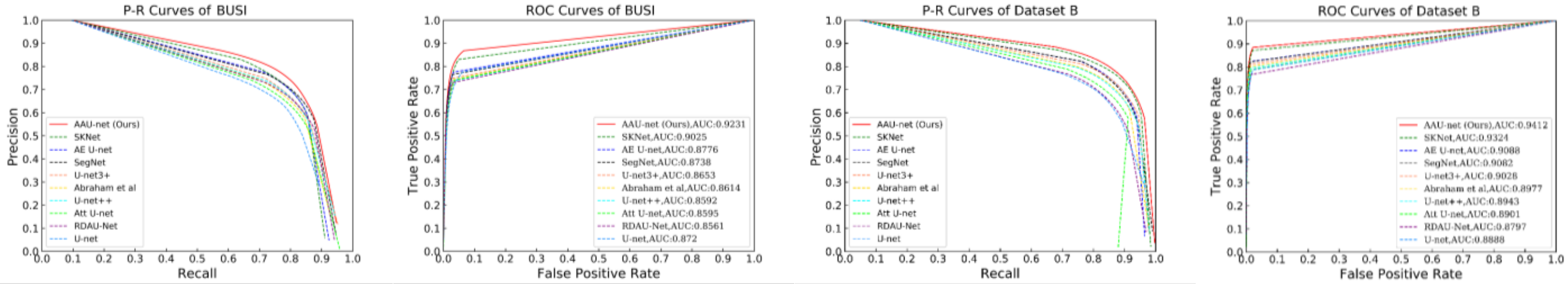

Fig. 4. P-R and ROC curves of different segmentation methods on BUSI and Dataset B.

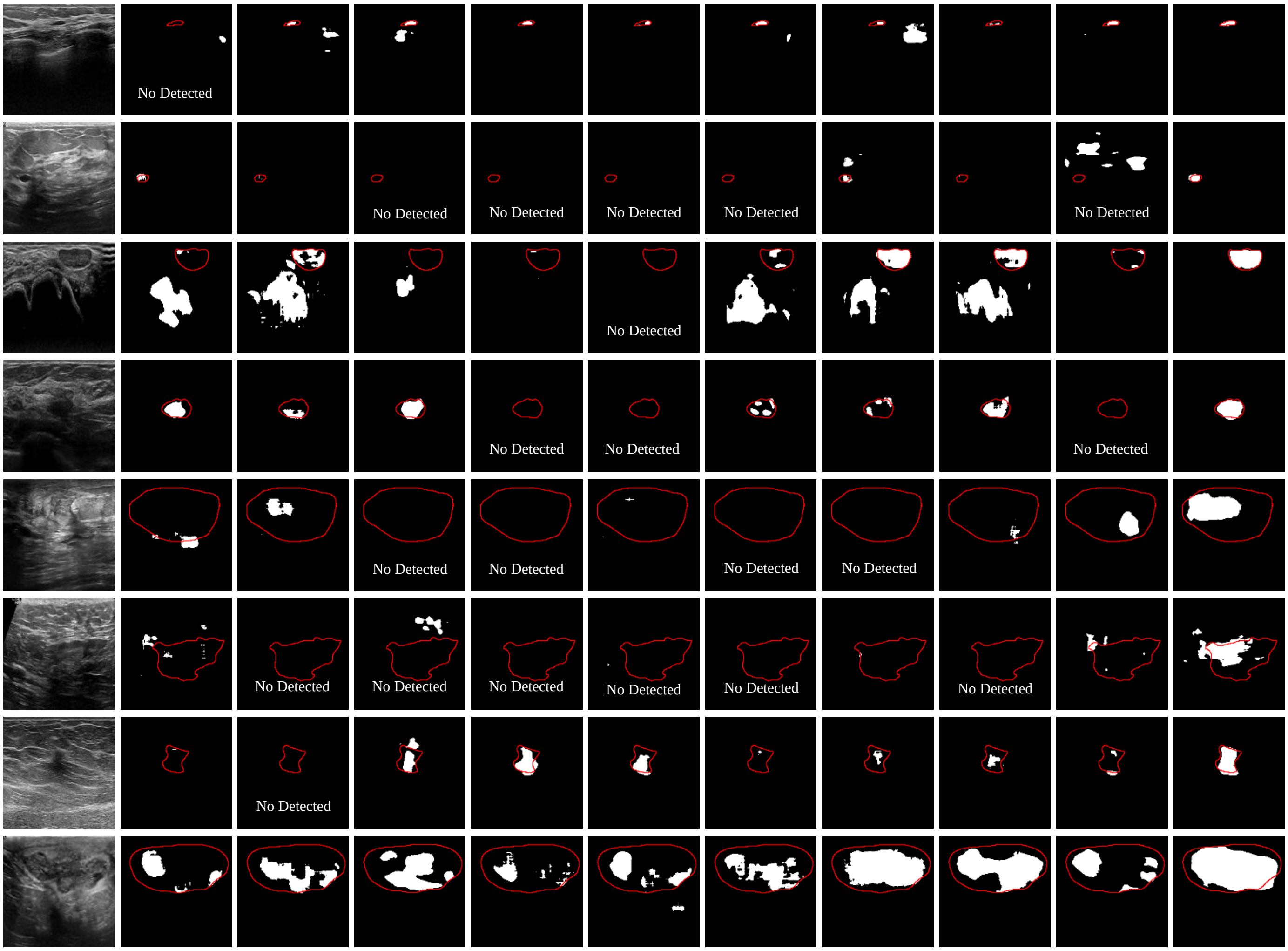

Fig. 5. Segmentation results of different methods on BUSI and Dataset B. From left to right are images of Input, U-net, Att U-net, RDAU-Net, U-net++, Abraham *et al.*, U-net3+, SegNet, AE U-net, SKNet and Ours. The red curve is the boundary of the breast lesion.

## C. Robustness Analysis

To further evaluate the robustness of our method, we first analyze the segmentation performance of different methods on benign and malignant breast tumors. Then, we use STU as external validation data to evaluate the performance of different methods. Furthermore, we evaluate the impact of normal ultrasound images in the network segmentation performance. Finally, we compare with several state-of-the-art attention-based segmentation methods.

### 1) Robustness on Benign and Malignant Lesions

Compared with benign lesions, malignant lesions have irregular shapes and blurred borders. In addition, the intensity distribution is more heterogeneous in malignant lesions compared to benign lesions. We conduct comparative experiments on benign and malignant BUS images of BUSI to evaluate the robustness of the network to segment malignant and benign lesions. We perform four-fold cross-validation on benign images and three-fold cross-validation on malignant images. Table V presents the segmentation results of different methods on malignant and benign breast lesions. Obviously, our method achieves higher scores in the segmentation of benign and malignant lesions. Moreover, the p-value compared to the second results indicates that our method has a significant improvement in segmentation accuracy. In Fig. 6 we draw the ROC curves of different segmentation methods on benign and

TABLE V

THE SEGMENTATION RESULTS (MEAN ± STD) OF BENIGN AND MALIGNANT LESIONS IN BUSI BY DIFFERENT METHODS. THE BEST RESULTS ARE MARKED WITH BOLD TEXTS. ASTERISKS INDICATE THAT THE DIFFERENCE BETWEEN OUR METHOD AND THE COMPETING METHOD IS SIGNIFICANT USING A PAIRED STUDENT'S T-TEST. (*: P < 0.05).

| | Method | U-net | Att U-net | RDAU-Net | U-net3+ | Abraham *et al.* | U-net++ | SegNet | AE U-net | SKNet | Ours |
|---|---|---|---|---|---|---|---|---|---|---|---|
| Benign | Jaccard | 61.53±3.98 | 65.03±2.05 | 64.70±2.17 | 67.63±1.86 | 66.74±2.10 | 68.25±2.75 | 67.89±3.31 | 67.89±1.96 | 69.91±2.11* | **73.33±2.09** |
| | Precision | 74.97±2.80 | 75.24±1.68 | 72.54±1.57 | 75.58±2.88 | 76.74±2.94 | 75.93±3.66 | 76.96±3.11 | 77.17±3.63 | 79.15±2.05* | **82.70±2.90** |
| | Recall | 73.97±5.81 | 79.44±2.84 | 79.36±0.98 | 81.07±1.27 | 79.97±1.64 | 81.58±1.09* | 79.57±2.21 | 80.54±1.25 | 81.54±2.17 | **83.14±0.87** |
| | Specificity | 97.72±0.59 | 97.68±0.62 | 97.79±0.28 | 97.72±0.58 | 97.75±0.60 | 97.74±0.62 | 97.98±0.46 | 97.95±0.60 | 98.06±0.52* | **98.39±0.47** |
| | Dice | 70.49±3.23 | 73.30±2.00 | 72.70±1.62 | 75.07±2.10 | 74.82±2.26 | 75.56±2.79 | 75.47±2.91 | 75.77±1.82 | 77.88±2.98* | **80.88±2.06** |
| Malignant | Jaccard | 51.11±2.62 | 51.12±2.35 | 51.63±1.62 | 54.77±3.55 | 54.12±2.96 | 54.03±3.03 | 54.89±1.78 | 55.38±1.77 | 57.06±2.42* | **60.60±1.70** |
| | Precision | 64.96±2.55 | 61.62±0.97 | 60.85±5.01 | 65.78±2.66 | 67.46±3.40 | 65.50±2.94 | 63.79±2.65 | 67.87±3.81 | 69.59±4.20* | **72.62±3.13** |
| | Recall | 68.86±4.27 | 72.57±2.17 | 71.89±2.55 | 74.38±3.21 | 72.36±5.05 | 73.43±2.10 | 76.25±4.02* | 72.88±3.47 | 73.58±6.75 | **76.63±5.66** |
| | Specificity | 93.63±1.28 | 93.12±1.00 | 93.47±1.45 | 93.82±1.06 | 93.94±1.25 | 93.73±1.31 | 94.00±1.14 | 94.43±1.33 | 94.65±1.49* | **95.11±1.27** |
| | Dice | 63.47±2.38 | 62.95±2.14 | 62.44±2.21 | 66.19±3.37 | 65.77±2.58 | 65.52±2.75 | 65.90±1.97 | 66.50±1.52 | 68.19±2.28* | **71.54±1.74** |

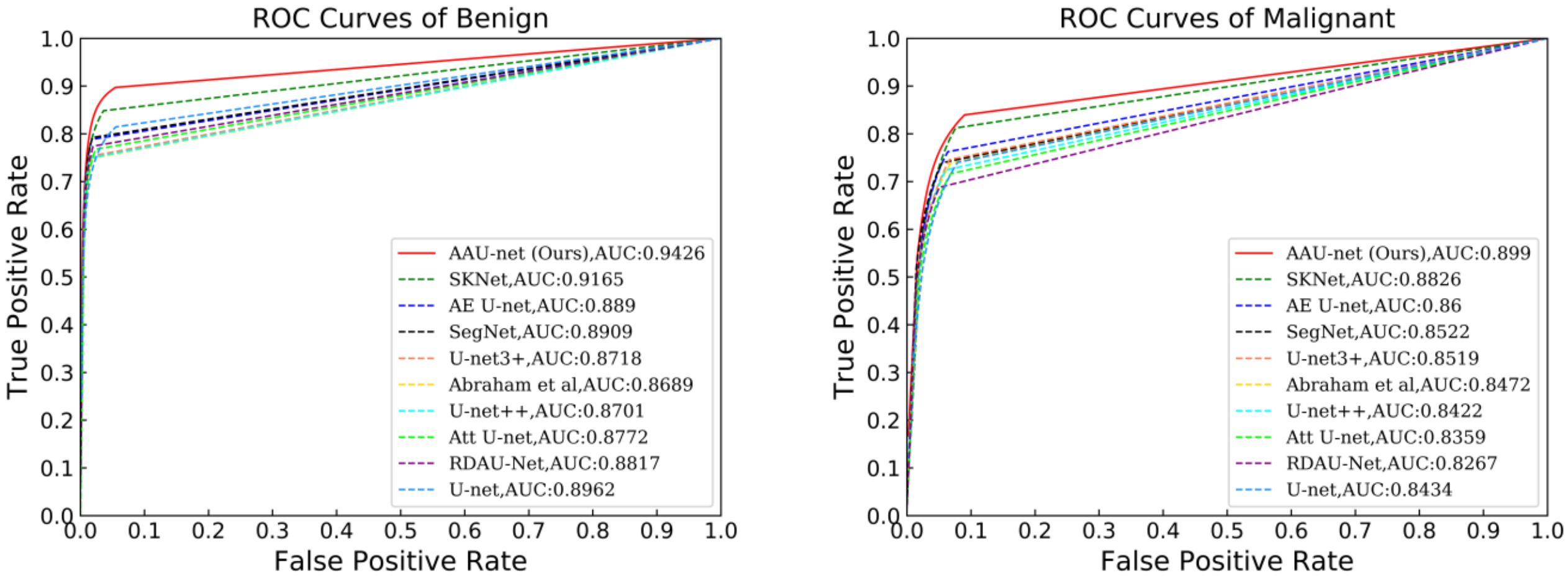

Fig. 6. ROC curves of different segmentation methods on benign and malignant breast lesions.

TABLE VI

SEGMENTATION RESULTS (MEAN ± STD) OF DIFFERENT COMPETING METHODS ON THE EXTERNAL VALIDATION DATASET STU. THE BEST RESULTS ARE MARKED WITH BOLD TEXTS. ASTERISKS INDICATE THAT THE DIFFERENCE BETWEEN OUR METHOD AND THE COMPETING METHOD IS SIGNIFICANT USING A PAIRED STUDENT'S T-TEST. (*: P < 0.05).

| | Method | Att U-net | U-net | U-net++ | RDAU-Net | Abraham *et al.* | U-net3+ | SegNet | AE U-net | SKNet | Ours |
|---|---|---|---|---|---|---|---|---|---|---|---|
| STU on Dataset B | Jaccard | 52.65±2.29 | 58.90±3.75 | 59.18±4.21 | 60.11±3.02 | 58.11±3.24 | 61.51±1.78 | 62.70±3.09 | 62.46±2.44 | 66.94±3.15* | **68.99±3.29** |
| | Precision | 59.26±2.86 | 66.27±5.46 | 64.86±5.36 | 63.49±2.86 | 66.05±3.69 | 67.12±2.22 | 66.57±3.05 | 66.79±2.95 | 71.99±4.08* | **74.91±3.18** |
| | Recall | 86.35±1.29 | 86.88±1.60 | 89.67±1.59 | 91.37±1.07 | 86.17±0.99 | 89.96±1.01 | 91.36±0.38 | 91.24±0.42 | 91.44±1.08* | **92.12±0.75** |
| | Specificity | 93.41±0.41 | 94.54±0.74 | 94.33±0.83 | 94.70±0.45 | 94.35±0.47 | 94.49±0.31 | 95.04±0.49 | 95.00±0.39 | 95.40±0.51* | **95.94±0.71** |
| | Dice | 65.19±2.73 | 71.41±3.67 | 70.70±4.19 | 72.40±2.74 | 70.32±3.01 | 72.96±1.69 | 73.50±3.62 | 74.41±2.30 | 78.29±3.05* | **80.23±2.60** |
| | HD | 85.01±2.04 | 74.05±11.63 | 64.61±6.56 | 52.55±3.04* | 83.38±10.10 | 72.10±10.03 | 65.58±2.11 | 62.86±5.00 | 57.37±9.69 | **45.50±3.20** |
| | ABD | 17.16±2.89 | 18.03±2.52 | 14.71±1.27 | 12.02±0.43 | 18.60±0.87 | 15.09±1.25 | 11.33±0.74 | 13.62±0.94 | 10.73±1.84* | **9.62±1.10** |
| | ASSD | 3.98±1.28 | 5.51±1.45 | 3.28±0.87 | 1.85±0.78 | 5.22±0.76 | 2.73±0.85 | 1.00±0.36* | 2.12±0.52 | 1.73±0.42 | **0.81±0.29** |

malignant breast lesions to further demonstrate the confidence level of our method. According to the ROC curves, we can clearly find that our method not only achieves convincing results on benign lesions, but also achieves the most competitive performance in the malignant lesion segmentation.

### 2) External Validation

Due to the differences between different sites, there are large differences between the collected data [20]. These differences can cause the model to perform well in the training dataset, but not perform well in the external data. To further evaluate the robustness of the proposed method in this paper, we use STU as external data to test the models trained on Dataset B by different methods. Compared with BUSI, Dataset B has a smaller number of samples. Therefore, comparative analysis on the smaller dataset (Dataset B) can better reflect the superiorities of different methods. The segmentation results of various methods on the external validation dataset STU are presented in Table VI. Our method still achieves the best results on eight evaluation metrics. This shows that the method proposed in this paper has better robustness compared to other methods, and is more suitable for breast lesions segmentation. In addition, the p-value based on Student's T-test also indicates the superiority of our method. SegNet, AE U-net and SKNet still obtain competitive results among the compared methods, which indicates that these three methods have certain potential in breast lesions segmentation. It is worth noting that the segmentation performance of Att U-net and U-net++ is degraded on STU. The occurrence of this phenomenon may be caused by their poor generalization ability. Fig. 7 shows the visual segmentation result of different methods on the external validation dataset STU. Visually, our method achieves the best segmentation results with fewer missed and false detections. From the first and fourth rows, it can be seen that the heterostructure affects the segmentation accuracy of various methods. Fortunately, our method is able to mitigate their perturbations. Overall, our method achieves the best segmentation results on external validation data STU. From the above analysis, it can be concluded that our method is insensitive to input data and has good generalization ability.

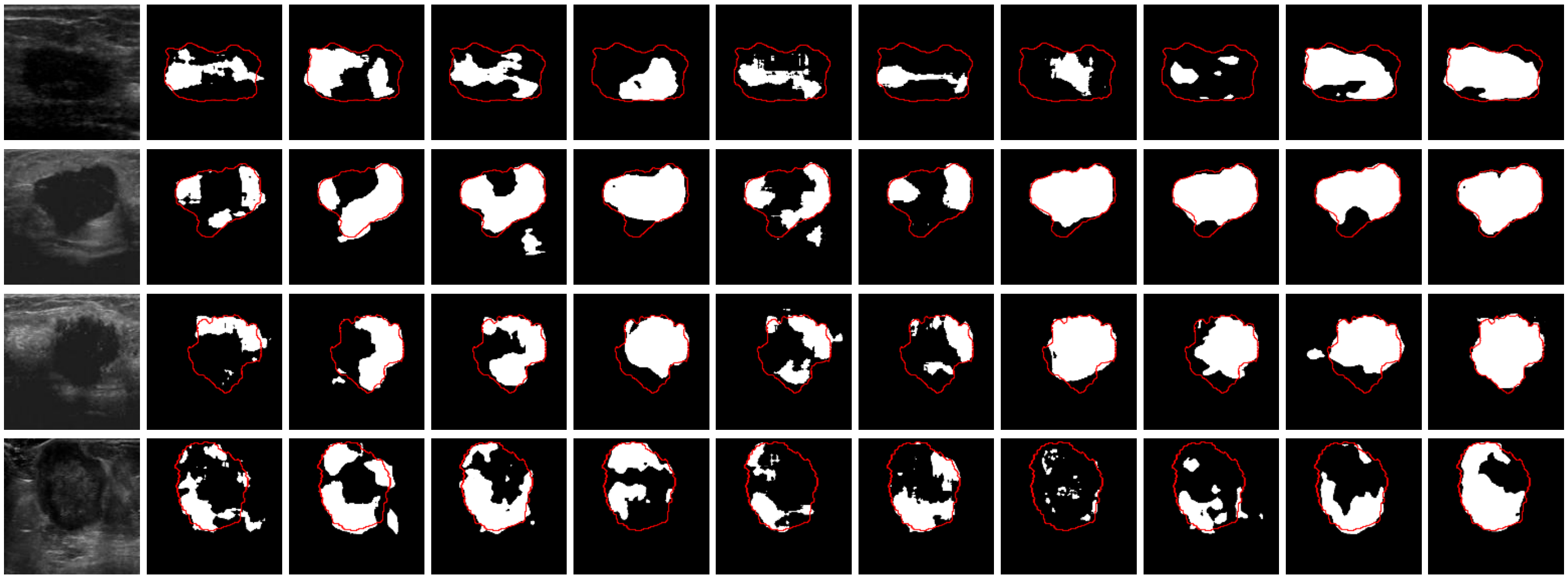

Fig. 7. Illustration of the segmentation results of different methods on STU. From left to right are images of Input, Att U-net, U-net, U-net++, RDAU-Net, Abraham *et al.*, U-net3+, SegNet, AE U-net, SKNet and Ours. The red curve is the boundary of the breast tumor.

TABLE VII
THE SEGMENTATION RESULTS (MEAN ± STD) OF DIFFERENT COMPETING METHODS ON BUSI WITH NORMAL IMAGES. THE BEST RESULTS ARE MARKED WITH BOLD TEXTS. ASTERISKS INDICATE THAT THE DIFFERENCE BETWEEN OUR METHOD AND THE COMPETING METHOD IS SIGNIFICANT USING A PAIRED STUDENT'S T-TEST. (*: $p < 0.05$).

| | Method | U-net | Att U-net | RDAU-Net | U-net++ | Abraham *et al.* | U-net3+ | SegNet | AE U-net | SKNet | Ours |
|---|---|---|---|---|---|---|---|---|---|---|---|
| BUSI (with normal images | Jaccard | 47.61±2.85 | 48.73±2.65 | 49.86±2.89 | 51.06±2.49 | 50.06±2.14 | 51.13±2.37 | 51.07±2.38 | 51.28±2.17 | 52.69±2.15* | **58.53±2.12** |
| | Precision | 61.23±2.35 | 60.85±2.54 | 59.68±2.57 | 63.05±2.81 | 63.55±2.73 | 62.64±3.05 | 62.85±3.10 | 63.84±3.97 | 66.84±2.83* | **68.16±3.52** |
| | Recall | 63.05±3.12 | 65.96±2.77 | 65.23±2.96 | 68.03±2.54 | 66.57±2.62 | 68.96±2.75* | 68.31±2.87 | 66.97±2.71 | 68.12±2.63 | **69.85±2.84** |
| | Specificity | 96.09±1.22 | 95.98±0.65 | 96.23±0.91 | 96.09±1.16 | 96.21±1.42 | 96.10±1.05 | 96.37±0.72 | 96.71±1.03 | 96.73±1.22* | **97.15±0.86** |
| | Dice | 56.02±1.98 | 58.39±2.47 | 56.87±2.43 | 59.41±2.27 | 59.06±2.45 | 59.32±2.31 | 59.85±2.21 | 60.49±2.97 | 61.52±2.58* | **65.83±2.55** |

### 3) Comparison on BUSI with Normal Images

The general purpose of breast lesions segmentation in the clinical usage is mainly for the lesion assessment, tracking the lesion change, and identifying distribution and seriousness of lesion. Therefore, people usually assume that the input ultrasound samples possess one or more lesions, and then conduct the breast lesion segmentation for clinical analysis. In this paper we conduct some new comparative experiments by introducing normal ultrasound images of BUSI. We perform four-fold cross-validation on BUSI with normal ultrasound images. Table VII presents the segmentation results of various methods on BUSI with normal ultrasound images. Compared with Table IV, the introduction of normal ultrasound images of BUSI severely affected the performance of the segmentation network. Similarly, existing work (GG-Net) also shows that the introduction of normal ultrasound images in BUSI is not beneficial for the segmentation of breast lesions [27]. From Table IV and Table VII, we can see that our method achieves the best segmentation performance on BUSI with and without normal ultrasound images. This indicates that our method can alleviate the perturbation of the segmentation results by surrounding tissues with similar intensity distributions to a certain extent. In addition, the p-value based on Student's T-test also indicates the superiority of our method.

### 4) Comparison with different Attention-Based Methods

In recent years, many attention-based methods have been proposed to improve the performance of networks, such as AGNet [40], SANet [41], ECA-Net [42], scSENet [43], SENet [44]. To further highlight the advantages of the hybrid adaptive attention module (HAAM), we compare with these five attention-based methods. In the experiment, we perform four-fold cross-validation on BUSI and Dataset B, respectively. The quantitative evaluation of the segmentation results of BUSI and Dataset B by different segmentation methods is presented in Table VIII. Our method still achieves the best segmentation performance on BUSI and Dataset B, as shown in Table VIII.

TABLE VIII
THE SEGMENTATION RESULTS (MEAN ± STD) OF DIFFERENT COMPETING METHODS ON BUSI AND DATASET B. THE BEST RESULTS ARE MARKED WITH BOLD TEXTS. ASTERISKS INDICATE THAT THE DIFFERENCE BETWEEN OUR METHOD AND THE COMPETING METHOD IS SIGNIFICANT USING A PAIRED STUDENT'S T-TEST. (*: $p < 0.05$).

| | BUSI | | | | | Dataset B | | | | |
|---|---|---|---|---|---|---|---|---|---|---|
| Methods | Jaccard | Precision | Recall | Specificity | Dice | Jaccard | Precision | Recall | Specificity | Dice |
| AGNet | 62.23±2.02 | 72.11±2.77 | 78.80±2.00 | 96.30±0.68 | 71.35±2.10 | 64.15±1.35* | 74.35±4.72* | 80.19±4.30* | 98.70±0.31* | 73.30±1.03* |
| SANet | 65.96±2.78 | 74.84±4.18 | 80.76±2.30* | 96.75±0.70 | 74.46±2.76 | 63.26±7.77 | 72.40±10.19 | 77.81±5.07 | 98.61±0.57 | 71.84±7.96 |
| SENet | 67.75±3.09 | 78.70±3.51 | 80.04±3.07 | 97.15±0.71 | 76.71±2.88 | 60.77±6.41 | 71.95±8.50 | 78.18±6.77 | 98.49±0.27 | 70.45±5.30 |
| scSENet | 67.68±2.28 | 78.95±2.73* | 79.58±1.14 | 97.26±0.48* | 76.67±2.20 | 62.17±5.03 | 71.31±5.44 | 79.16±5.79 | 98.52±0.22 | 71.30±4.24 |
| ECA-Net | 68.17±2.21* | 78.59±2.77 | 80.73±1.91 | 97.20±0.53 | 77.10±2.17* | 63.23±5.26 | 72.65±4.19 | 78.53±4.99 | 98.62±0.26 | 72.09±4.74 |
| Ours | **68.82±0.44** | **79.61±1.07** | **81.10±0.52** | **97.57±0.24** | **77.51±0.68** | **69.10±2.98** | **78.83±2.40** | **82.22±3.84** | **98.82±0.35** | **78.14±2.41** |

From Table VIII, we can see that the compared methods have inconsistent segmentation performance on BUSI and Dataset B. This shows that these methods are more sensitive and less robust to different input data. Furthermore, the p-value based on Student's T-test indicates that our method is significantly different from these comparative methods. The comparison with attention-based methods further shows that our method has better robustness and generalization ability.

## V. DISCUSSION

In this study, we propose a novel adaptive attention U-net (AAU-net) to alleviate the challenge of breast lesions segmentation. To evaluate the effectiveness of network components and parameters, we first perform the ablation study. According to the experimental results in Table II and Table III, we can clearly see that the settings of our network components and parameters enable the network to achieve the best performance in the breast lesion segmentation.

According to the comparative experimental results of state-of-the-art segmentation methods, we can draw several conclusions. In general, the U-net-based variant network (such as U-net++ and U-net3+) achieves better segmentation results than the original U-net, which indicates that the use of skip-connection operations to fuse low-level features in the encoding stage with high-level features in the decoding stage is beneficial for the segmentation of breast lesions. By analyzing the segmentation results of the U-shaped network with the attention mechanism (such as RDAU-Net), it can be concluded that the introduction of the attention mechanism can also improve the segmentation performance of the network. Compared with the results of AE U-net and SKNet, the optimized attention mechanism can further improve the performance of the network to segment breast lesions. Compared with U-net, Att U-net has a poor segmentation result on BUSI, but it has a better segmentation result on Dataset B. This shows that the use of this attention mechanism will increase the sensitivity of the network to breast ultrasound images. From the segmentation results of the method proposed by Abraham et al., it can be seen that introducing the strategy of deep supervision and multi-scale inputs into Att U-net can help improve the segmentation accuracy of breast lesions.

Based on the segmentation results of SegNet, it can be seen that using the location information of the features in the U-shaped network can achieve better results than most segmentation methods. Generally speaking, SKNet achieves good performance on breast lesions segmentation among the compared methods. Compared with SKNet, the larger receptive field and spatial self-attention block introduced by our method can effectively improve the performance of the network in segmenting breast lesions. Based on the visual segmentation results shown in Fig. 5, we can summarize four key points. According to the segmentation results of the first and second rows in Fig. 5, we can see that various methods have some missed detection for small breast tumors, and even fail to detect breast tumors. From the segmentation results from the first row to fourth row in Fig. 5, we can conclude that the surrounding tissue (background) with similar intensity distribution can cause serious missed detections and false detections of breast lesions. Severe heterogeneity makes breast lesions undetectable by various methods as shown in the fifth and sixth rows in Fig. 5. Furthermore, accurate tumor contours cannot be captured from blurred or cascaded BUS images as shown in the last two rows of images in Fig. 5. Although our method still suffers from false detections and missed detections, it achieves significant improvements compared to other methods.

The experimental results of the robustness analysis not only demonstrate the good generalization ability of our network, but also further highlight the advantages of the hybrid adaptive attention module (HAAM). From the robustness experiments on benign and malignant lesion segmentation, we can see that the segmentation performance of various methods is relatively stable. Although the advantage of our method is reduced, it still achieves better segmentation performance than other compared methods. This shows that our method can well adapt to different data inputs. The experimental results of external validation also further prove that our network has good generalization ability. According to the experimental results with attention-based segmentation methods, it can be concluded that their segmentation performances on BUSI and Dataset B are significantly different, which indicates that they are sensitive to different input data. The method proposed in this paper achieves consistently good segmentation results on BUSI and Dataset B. Compared with existing attention modules, our hybrid adaptive attention module (HAAM) can help the network learn the more generic representation of breast lesions from ultrasound images. From the above analysis, it can be concluded that our hybrid adaptive attention module (HAAM) outperforms existing attention models in the breast lesion segmentation.

Although the method proposed in this paper has achieved good performance on breast lesions segmentation, it can be seen from Fig. 5, Fig. 7 and Fig. 8 that our method still has some shortcomings. (1) For more complex BUS images segmentation, our method still needs to be further optimized to reduce the false detection and missed detection rate. (2) How to obtain accurate object contours is still a challenging task. (3) The similar intensity distribution of surrounding tissues seriously affects the segmentation accuracy of breast lesions. To alleviate the above challenges, we will introduce boundary constraints to further improve the segmentation performance of the network. In addition, designing a reasonable data augmentation algorithm to expand the sample space is also our research direction.

## VI. Conclusion

To better address the challenge of breast tumor segmentation, we design a novel hybrid adaptive attention module (HAAM) and use it to construct an adaptive attention U-net (AAU-net) for breast lesions segmentation. The hybrid adaptive attention module can guide the network to adaptively select more robust representation in both channel and space dimensions to cope with more complex breast lesions segmentation. Extensive experiments (comparative experiments, robustness analysis and external validation) with several state-of-the-art deep learning segmentation methods demonstrate that our method has better performance on breast lesions segmentation. The source code is publicly available on https://github.com/CGPxy/AAU-net